\documentclass[11pt]{article}
\usepackage[margin=1in]{geometry}
\usepackage{authblk}
\usepackage{natbib}
\usepackage[utf8]{inputenc}
\usepackage{textcomp}
\usepackage{graphicx}
\usepackage{amsmath}
\usepackage{bm}
\usepackage[version=4]{mhchem}
\usepackage{siunitx}
\usepackage{multirow}
\usepackage{derivative}
\usepackage{physics}
\usepackage{longtable,tabularx}
\usepackage{setspace}
\usepackage[pagewise]{lineno}
\usepackage{booktabs}
\usepackage{tabularray}
\UseTblrLibrary{booktabs,siunitx}
\usepackage{subcaption}
\usepackage[toc,page]{appendix}

\usepackage[dvipsnames]{xcolor}
\usepackage{hyperref}
\hypersetup{
colorlinks=true
,linkcolor=BrickRed
,citecolor=OliveGreen
,filecolor=Mulberry
,urlcolor=NavyBlue
,menucolor=BrickRed
,runcolor=Mulberry
,linkbordercolor=BrickRed
,citebordercolor=OliveGreen
,filebordercolor=Mulberry
,urlbordercolor=NavyBlue
,menubordercolor=BrickRed
,runbordercolor=Mulberry
,pdftitle={}
,pdfpagemode=FullScreen
}

\usepackage{tikz}
\def\checkmark{\tikz\fill[scale=0.4](0,.35) -- (.25,0) -- (1,.7) -- (.25,.15) -- cycle;} 

\usepackage{xargs}
\usepackage[colorinlistoftodos,prependcaption,textsize=tiny,disable]{todonotes}
\newcommandx{\unsure}[2][1=]{\todo[linecolor=red,backgroundcolor=red!25,bordercolor=red,#1]{#2}}
\newcommandx{\change}[2][1=]{\todo[linecolor=blue,backgroundcolor=blue!25,bordercolor=blue,#1]{#2}}
\newcommandx{\info}[2][1=]{\todo[linecolor=teal,backgroundcolor=teal!25,bordercolor=teal,#1]{#2}}
\newcommandx{\improvement}[2][1=]{\todo[linecolor=magenta,backgroundcolor=magenta!25,bordercolor=magenta,#1]{#2}}
\newcommandx{\thiswillnotshow}[2][1=]{\todo[disable,#1]{#2}}

\newcounter{mcounter}

\newcommand{\incfig}{\centering\includegraphics}
\newcommand{\mach}{M_{\infty}}

\usepackage{microtype}
\usepackage[capitalize]{cleveref}

\title{Compressibility Effects on Leading-Edge Dynamic Stall Criteria at High Reynolds Number}

\author{Sarasija Sudharsan and Anupam Sharma}
\affil{Iowa State University, Ames, IA, 50011}

\begin{document}
\maketitle
\onehalfspacing

\begin{abstract}
This study examines the applicability of two leading-edge dynamic stall criteria, namely, the maximum magnitudes of the leading-edge suction parameter (LESP) and the boundary enstrophy flux (BEF), in a moderately compressible flow regime. 
While previously shown to predict stall onset ahead of dynamic stall vortex (DSV) formation in incompressible and mildly compressible regimes, these criteria are assessed here at a Reynolds number of $1 \times 10^6$ and freestream Mach numbers between 0.3 and 0.5. 
Unsteady RANS simulations indicate that DSV formation occurs in close temporal proximity to the attainment of the stall criteria. 
However, at the highest Mach number considered, stronger shock interaction effects with the shear layer leads to DSV formation prior to the criteria being reached, reducing their predictive accuracy. 
These findings suggest that while the criteria remain effective at lower Mach numbers, their definitions require modification in compressible regimes where strong shock interactions significantly influence the stall process.
\end{abstract}

\listoftodos

\section{Introduction}

Dynamic stall is the unsteady stall phenomenon that occurs over aerodynamic surfaces undergoing large amplitude transient motion or unsteady maneuvers, for example, over wind turbine or helicopter rotor blades.
The modeling and control aspects of dynamic stall have interested aerodynamicists for several decades, since the large, unsteady aerodynamic forces and moments incurred could potentially cause catastrophic structural failure~\citep{McCroskey1981, Corke2015}. 
Stall control efforts are most effective before the formation of the dynamic stall vortex (DSV)~\citep{Chandrasekhara2007}, a characteristic feature of `deep' dynamic stall.
Therefore, characterizing stall onset is of crucial importance for control efforts to be deployed in a timely manner.
Various criteria for dynamic stall onset based on the unsteady aerodynamic coefficients have been explored to formulate first-order, semi-empirical, dynamic stall models~\citep{Leishman1989, Sheng2005}.
However, the leading edge suction parameter ($LESP$)~\citep{Ramesh2014,Narsipur2020} and the boundary enstrophy flux ($BEF$)~\citep{Sudharsan2022} have been used to narrow down the identification of stall onset to a finer degree in time~\citep{Sudharsan2023, Sudharsan2024}.
While these two criteria have previously been applied to a lower $Re$ of $2 \times 10^5$ in the compressible regime~\citep{sudharsan2023effects}, the current work extends their application to compressible flows at $Re = 1 \times 10^6$. 
Some comparisons with LES results from \citet{benton2020effects} at this high $Re$ are also presented.

Prior studies, e.g. \citet{McCroskey1981}, indicate that compressibility effects become significant in unsteady flows at freestream Mach numbers as low as $0.2$.
The fundamental mechanisms of dynamic stall are influenced by the degree of compressibility, which can be categorized into mildly, moderately, and highly compressible regimes.
In the \emph{mildly-compressible regime}, the flow remains sub-critical, i.e., the maximum local flow Mach number over the airfoil, $M_{\rm loc} < 1$. 
Compressibility promotes adverse-pressure-gradient (APG)-induced leading-edge stall and reduces dynamic-stall delay and dynamic lift overshoot~\citep{CARR1996523}.
In the \emph{moderately-compressible regime}, the maximum local flow speed exceeds the sonic speed, but it does not induce shocks strong enough to cause significant boundary layer separation. 
However, the subsonic flow inside the laminar separation bubble (LSB) can be fully or partially surrounded by locally-supersonic flow.
The stall mechanism here is due to the interaction between the LSB and small wave-like disturbances in the surrounding flow~\citep{chandrasekhara1998competing}, which leads to LSB bursting. 
This fundamentally differs from the APG-induced LSB bursting in the mildly compressible regime. 
In the \emph{highly-compressible regime}, locally-supersonic flow can produce a complex shock structure (lambda shocks, for instance) near the blade leading edge. 
Compressibility forces premature boundary layer separation, and the type of shock dictates the trailing-edge boundary layer characteristics - weak shocks thicken the boundary layer, modestly strong shocks cause local boundary-layer separation followed by reattachment (separation bubble), and strong shocks cause massive trailing-edge flow separation and induce stall~\citep{chandrasekhara1998competing}. 
In shock-induced stall, the flow near the leading edge can remain unaffected even after the boundary layer over the rest of the airfoil is separated and a substantial loss of lift has occurred, see e.g., \citet{dadone1977one}. 
These observations are in stark contrast to those observed in the incompressible regime where APG is the primary stall mechanism. 
For $Re<10^6$, APG leads to LSB burst and triggers stall, while at higher $Re$, APG causes flow reversal which grows from the trailing to the leading edge, followed by dynamic stall vortex (DSV) shedding from the leading-edge region, resulting in stall.

Compressibility effects in dynamic stall has gained recent interest.
\citet{lee2024numerical} presented LES results for an NACA 0012 wing section undergoing unsteady stall during oscialltory pitching motion. They investigated three $Re$ ($100k,~200k,$ and $400k$) at $M_\infty = 0.1$ and $0.4$.
The focus of the study was to investigate flow characteristics and effects of compressibility on stall and its onset.
\citet{benton2020effects} analyzed the onset of dynamic stall for a pitching NACA 0012 airfoil at $Re = 1\times 10^6$ and $M_\infty = 0.1,~0.2,~0.3$ and $0.4$ using their LES results. 
They observed that compressibility leads to earlier bursting of the LSB, hence, early stall.
This was attributed to fundamental changes in the transition mechanism in the LSB due to compressibility.
In an earlier conference paper~\citep{sudharsan2023effects}, we investigated how compressibility influences these criteria at a lower Reynolds number ($Re = 2\times10^5$) over a freestream Mach number range of $M_\infty = 0.1\text{–}0.5$. 
In that regime, shocks did not significantly affect stall onset; instead, stall was driven by an increasing adverse pressure gradient (APG) near the leading edge. 
Both criteria were reached before DSV formation, which occurred entirely in subsonic flow. 
The reason is attributed to the criteria being reached coinciding with the collapse in leading-edge suction due to increasing APG downstream of the suction peak.
This collapse of leading-edge suction triggers LSB bursting and DSV formation.
Since compressibility amplifies APG-induced stall, we anticipate that these criteria remain directly applicable in such a ``mildly-compressible'' regime. 
In moderately-compressible conditions, however, shock-shear-layer interactions and subsonic near-wall flow can lead to shock-induced separation downstream of the leading edge—precisely where $LESP$ and $BEF$ are evaluated.

In the present work, we assess the effectiveness of two state-of-the-art stall indicators—$\max(\abs{BEF})$ and $\max(LESP)$—in predicting the onset of compressible dynamic stall at a high Reynolds number. 
We extend our study to a higher $Re$, namely, $1\times10^6$, over a Mach number range ($M_{\infty} = 0.3\text{–}0.5$). 
We compare uRANS predictions from the present work with LES results reported by ~\citet{benton2020effects} for $M_\infty = 0.3\text{–}0.4$.
We also introduce uRANS data at $M_\infty = 0.5$, where shock effects on stall become more pronounced.

\section{Methods}
\label{sec:methods}

\subsection{Solver, dataset, and grid details}
Our analysis utilizes numerical results obtained using wall-resolved LES and unsteady Reynolds-averaged Navier-Stokes (uRANS) simulations.
The LES results are from \citet{benton2020effects} and \citet{Sharma2019}, which were obtained using the compressible flow solver FDL3DI~\citep{gaitonde1998high}. 
Details on FDL3DI can be found in \citet{gaitonde1998high} and \citet{Visbal2002}.

The uRANS equations are solved using the open-source code Stanford Unstructured (SU2)~\citep{SU2}. 
The compressible Navier-Stokes equations written in strong conservation form are solved by discretization using a finite volume method, with an implicit, second-order, dual-time-stepping~\citep{jameson1991time} approach for time integration. 
Convective fluxes are calculated using the low-dissipation Low Mach Roe ($L^2$ Roe)~\citep{osswald2016l2roe} model for the two lower Mach numbers, while the classic Roe scheme~\citep{roe1981approximate} is used for the highest Mach number simulated. 
The two-equation SST $k-\omega$ turbulence model (version V2003m)~\citep{Menter2003}, with a freestream turbulence intensity level of 1\%, is used for closure.

Transitional effects are important to accurately capture stall onset at the $Re$ considered here. 
uRANS predicts delayed stall onset compared to LES (see \cref{sec:results}), which is attributed to the absence of a transition model in the former case.
However, the focus of the current work is on evaluating the stall-onset criteria for given flow solutions.
If the criteria predict stall onset around the time where the given flow solution exhibits stall, then the criteria are deemed effective.

\Cref{tab:datasets} lists the datasets used in the present work.
Flow over an NACA 0012 airfoil undergoing a constant-rate, pitch-up motion, pivoted about the quarter-chord point is simulated.
The unsteady motion occurs at a nondimensional pitch rate, $\Omega^* = \Omega_0 c/U_{\infty} = 0.05$, where $\Omega_0$ is the angular velocity of the pitching motion.
Simulations are carried out at freestream Mach numbers, $M_{\infty}$, of \qtylist{0.3;0.4;0.5}{} for $Re$ $1 \times 10^6$.
Stationary airfoil (static) simulations at $\alpha = \ang{4}$ are first carried out until a statistically stationary solution (for LES) and a converged solution (for RANS) is obtained.
A smooth, hyperbolic-tangent-based ramp function is prescribed to transition from the static simulation to the final nondimensional pitch rate. 

\begin{table}%[htpb]
	\centering
	\caption{Datasets used in the present work.
    In all cases, an NACA 0012 airfoil is pitched up at a constant rate about the quarter-chord point at a nondimensional pitch rate of $\Omega^* = 0.05$.}
    \label{tab:datasets}
	\begin{tabular}{c c c c}
		$Re$ & $M_\infty$ & uRANS & LES \\
		\toprule
		\multirow{3}{*}
        {$1 \times 10^6$} & $0.3$ & \checkmark & \checkmark \\ 
		& $0.4$ & \checkmark & \checkmark \\ 
		& $0.5$ & \checkmark &  -- \\ 
		\bottomrule
	\end{tabular}
\end{table}

\begin{figure}[htpb]
    \hspace*{\fill}
	\incfig[width=0.32\linewidth]{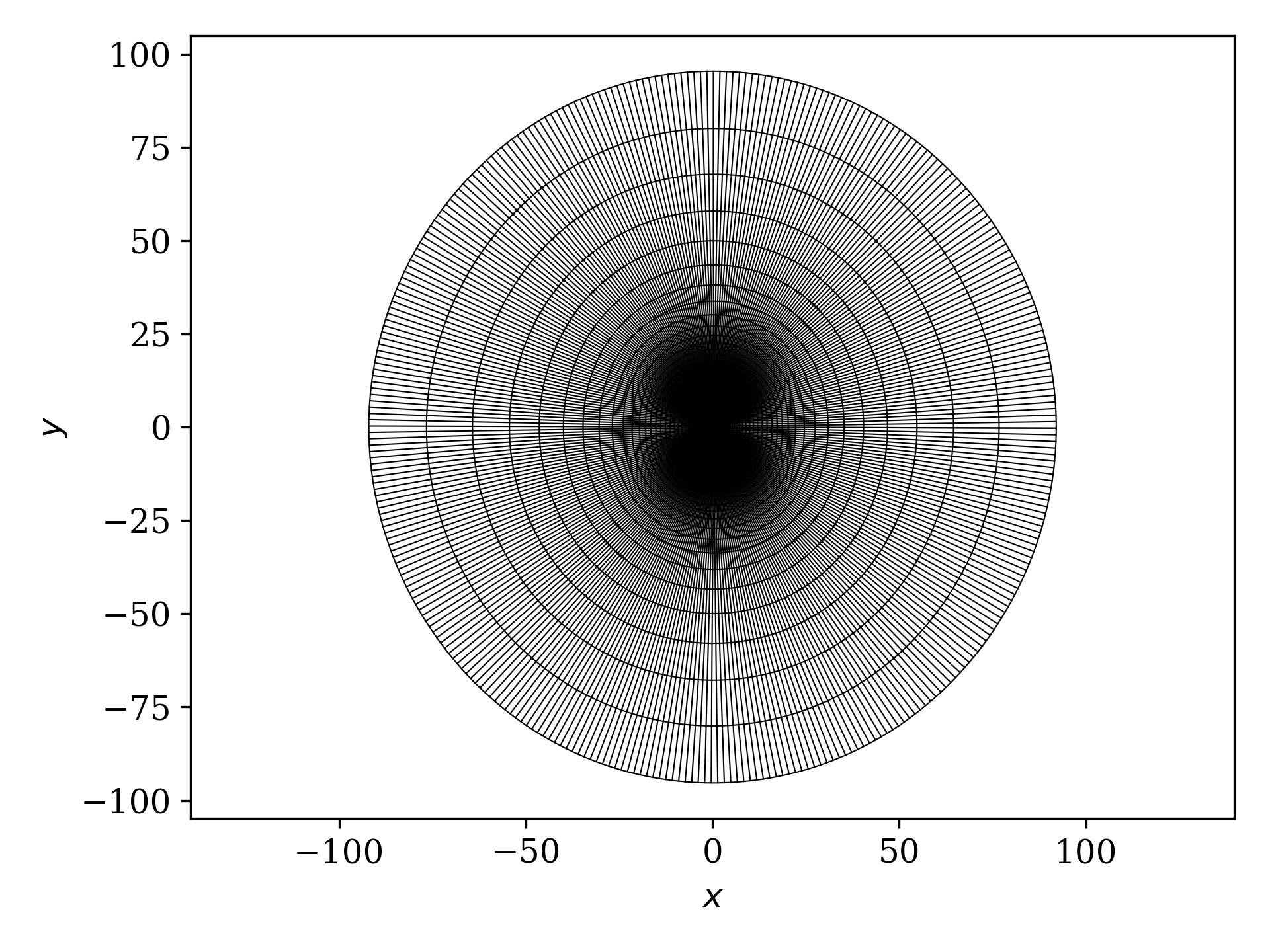}
    \hfill
	\incfig[width=0.32\linewidth]{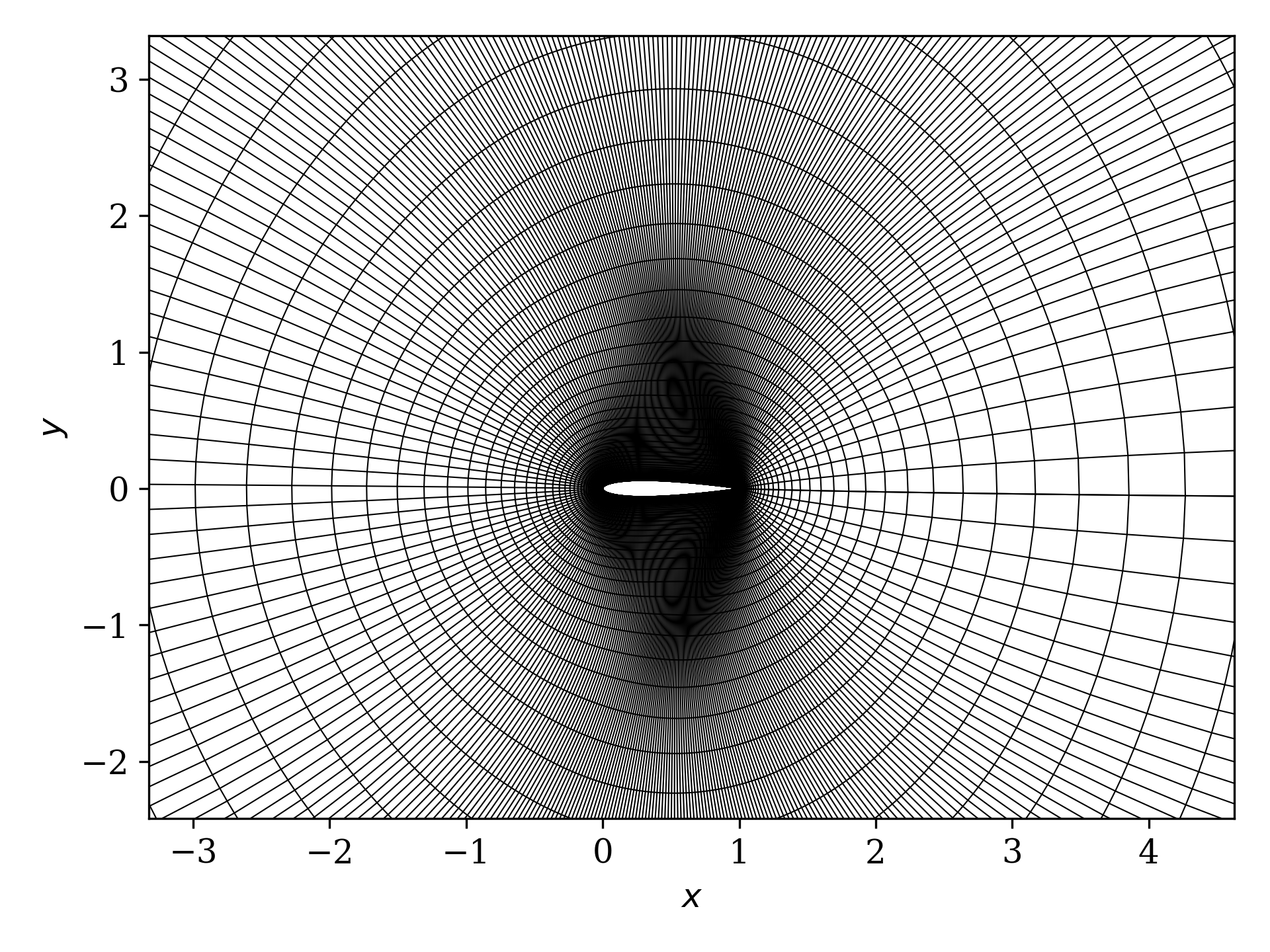}
    \hfill
	\incfig[width=0.32\linewidth]{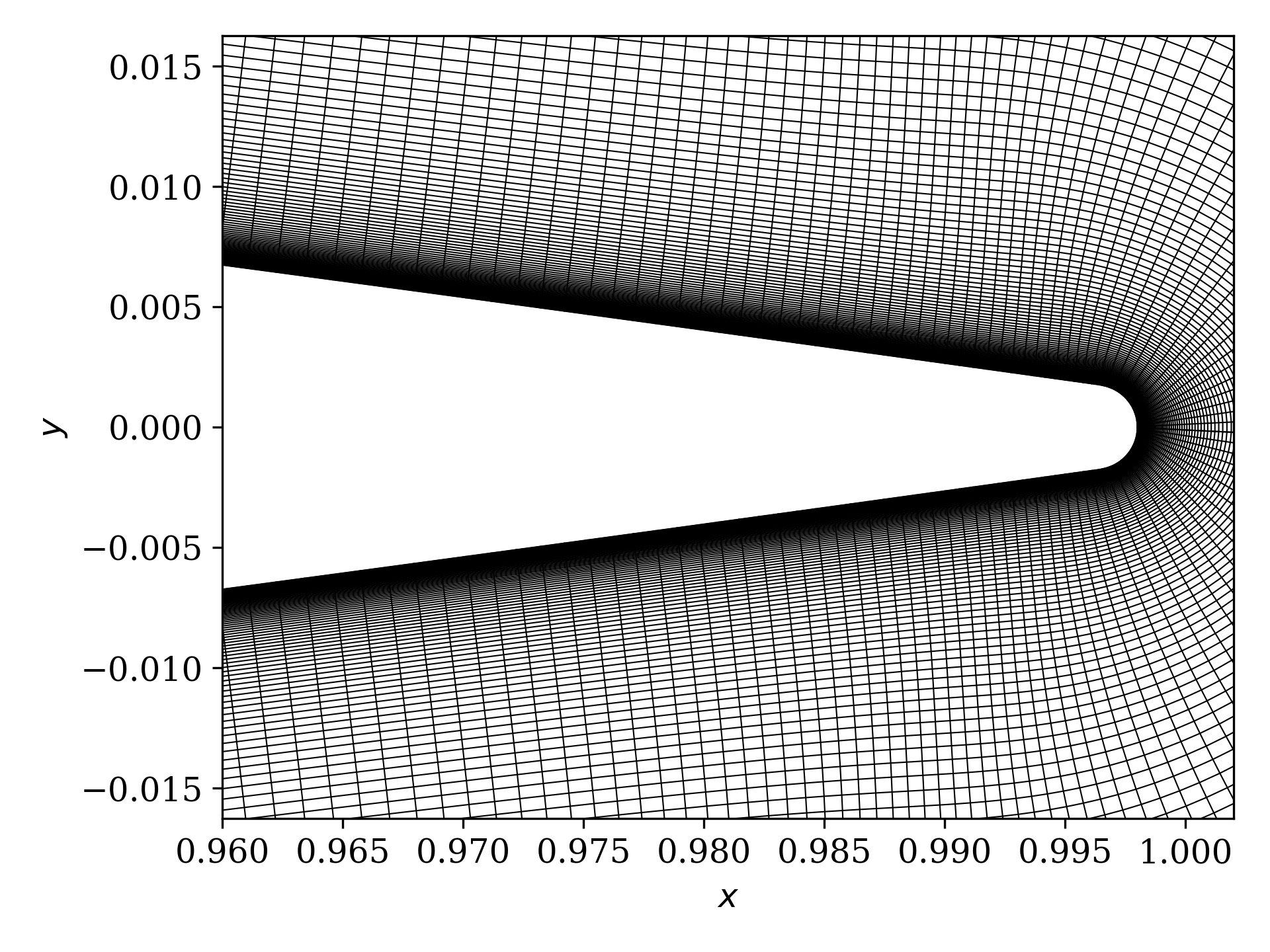}
    \hspace*{\fill}
	\caption{Grid used in the present study for uRANS simulations: full view (left), zoomed-in view (middle) and trailing-edge region (right). Every third point in the radial and circumferential directions are shown for clarity in the left and middle panels.
	\label{fig:grid}}
\end{figure}

The meshes used for the LES are described in \citet{benton2020effects}.
For the RANS simulations, a structured O-Mesh having $996$ and $288$ points in the circumferential and radial direction respectively, is used for all the simulations, with $y_+$ values less than $1$ over the entire airfoil surface.
\Cref{fig:grid} shows different views of the grid used for the uRANS simulations.
The grid is finalized based on comparisons with LES results~\citep{Sharma2019,Sudharsan2022} for $M_{\infty} = 0.1$.
A nondimensional time step, $ \Delta t^* = \Delta t U_{\infty}/c$, of $2 \times 10^4$ was used for time integration.

\subsection{Stall criteria definitions}
We first provide the definitions of the $BEF$ and $LESP$ parameters; detailed descriptions are available in Refs.~\citep{Ramesh2014,Narsipur2020,Sudharsan2022}.
For a 2D flow field, the $BEF$ is the flux of squared spanwise vorticity ($\omega$) at the wall. 
In \cref{eq:bef_def}, $\omega$ is normalized by $U_{\infty}/c$, where $U_{\infty}$ is the freestream velocity, and $c$, the airfoil chord.
The normal and tangential directions to the airfoil surface, $n$ and $s$, respectively, are normalized by $c$.
The integral is carried out from $x/c$ on the pressure side to $x/c$ on the suction side, with all quantities calculated in the airfoil frame of reference.
The inclusion of $Re_c$ is to enable expanding out the integrand, $|\omega^2|$, into a product of vorticity and boundary vorticity flux scaled by $Re_c$, which is equivalent to the favorable pressure gradient for small tangential acceleration.
Large contributions to the $BEF$ arise only from regions of high vorticity combined with large pressure gradients, which is primarily the leading-edge region. 
The $BEF$ is therefore nearly independent of integration length, as long the region very close to the leading edge (about 1\% chord based on prior studies~\citep{Sudharsan2022}) is included.
\begin{equation}
    BEF = \frac{1}{Re}\int_{(x/c)_p}^{(x/c)_s} \omega \pdv{\omega}{n} \,\dd s
    \label{eq:bef_def}
\end{equation}

$LESP$, given by \cref{eq:LESP_def}, is a measure of the camber-wise suction force ($F_{s,LE}$) near the leading edge, obtained by integrating the pressure coefficient $C_p$.
$q_{\infty}$ is the freestream dynamic pressure.
The integral to calculate the force is carried out from the maximum thickness point on the pressure side to that on the suction side.
\begin{equation}
    LESP = \sqrt{{C_{s,LE}}/{2 \pi}}, \quad {\rm where} \quad C_{s,LE} =  F_{s,LE}/q_{\infty}c
    \label{eq:LESP_def}
\end{equation}

\section{Results \& Discussion}
\label{sec:results}

We first discuss the results for $Re=1 \times 10^6$.
The LES computed flowfields for this $Re$ are described in \citet{benton2020effects} and hence only the uRANS results are presented here.
The LES results, along with uRANS, are utilized to evaluate the stall-onset criteria.

\subsection{Effect of compressibility on stall onset}
\begin{figure}[htpb!]
	\incfig[width=0.7\textwidth]{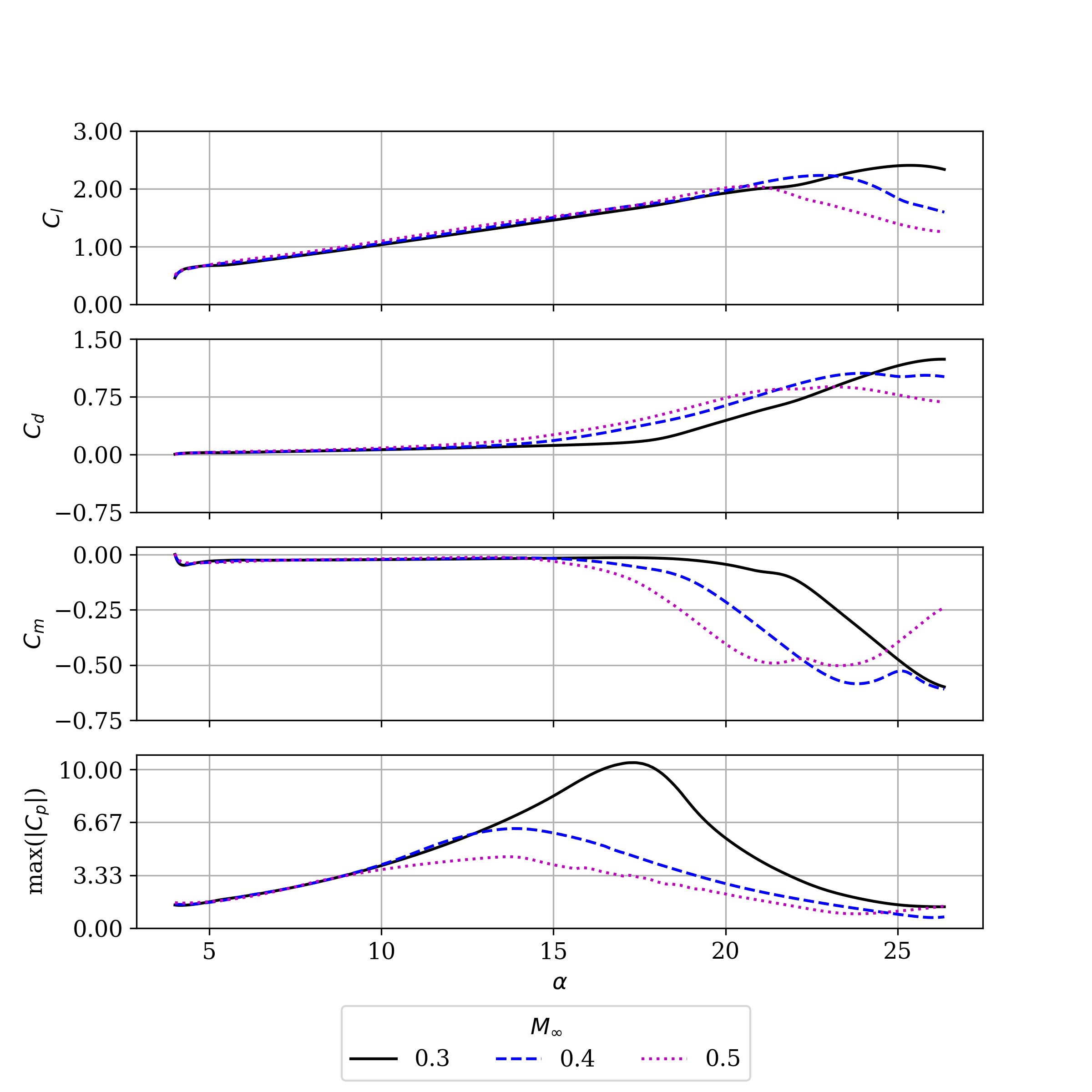}
    \caption{Variation in unsteady aerodynamic coefficients, $C_l$, $C_d$ and $C_m$, and maximum magnitude of $C_p$ near the leading edge as the airfoil pitches up, for the uRANS cases.
\label{fig:aerodyn_coeff_all_re1m}}
\end{figure}

Compressibility leads to earlier APG-induced, leading-edge stall, attributed to characteristic changes in the transition process near the leading edge of the airfoil~\citep{benton2020effects}.
Accordingly, we observe earlier occurrences of moment and lift stall (see \cref{fig:aerodyn_coeff_all_re1m}) as $\mach$ increases.
The top three panels in the figure show the variation of unsteady aerodynamic coefficients ($C_l, C_d, C_m$) with increasing $\alpha$ for different uRANS cases.
The bottom panel shows a point quantity, namely, the maximum magnitude of $C_p$ on the airfoil surface in the first 5\% chord near the leading edge, representing peak suction.
For a given $\mach$, peak suction collapse occurs earlier than lift stall; in the time between these two instances, shed vorticity close to the wall organizes into a coherent DSV inducing additional lift over the airfoil surface.
Therefore, a portion of the increase in  $C_l$ is attributed to vortex-induced lift due to DSV formation.
Moment stall occurs when the DSV convects past the aerodynamic center and contributes to a nose-down pitching moment.
This is observed as a marked divergence in the profile of $C_m$.
Moment stall occurs earlier than lift stall for all cases, since DSV formation continues to increase the lift over the airfoil surface while it remains attached to the leading-edge shear layer.
Therefore, the sequence of events goes as follows: peak suction collapse \textrightarrow moment stall \textrightarrow lift stall.
As $\mach$ increases, the magnitudes of the aerodynamic coefficients are lower and occur earlier, due to earlier breakdown of the leading-edge flow.
A progressively weaker DSV is formed as $M_{\infty}$ increases, resulting in lower vortex-induced lift contribution, and smaller divergences in $C_d$ and $C_m$.
The second minimum observed in $C_m$ around $\alpha=\ang{24}$ for $M_{\infty} = 0.4~\&~0.5$ (\cref{fig:aerodyn_coeff_all_re1m}) corresponds to a trailing-edge vortex that rolls up as the DSV is shed.
 
\begin{figure}[htpb]
    \centering
    \hspace*{\fill}
    \subcaptionbox{$C_l$\label{fig:cl_urans_les}}{\incfig[width=0.49\textwidth]{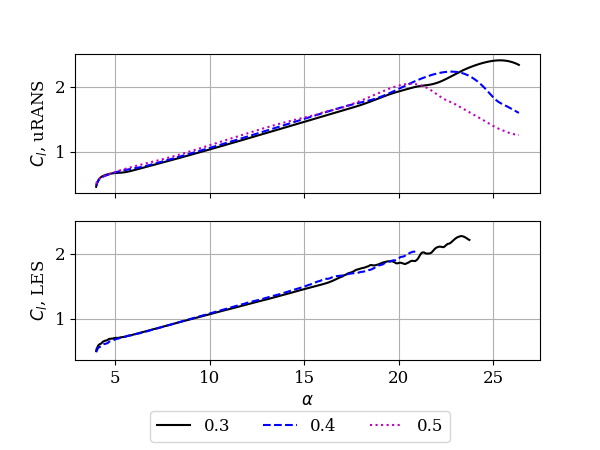}}
    \hfill
    \subcaptionbox{$C_m$\label{fig:cm_urans_les}}{\incfig[width=0.49\textwidth]{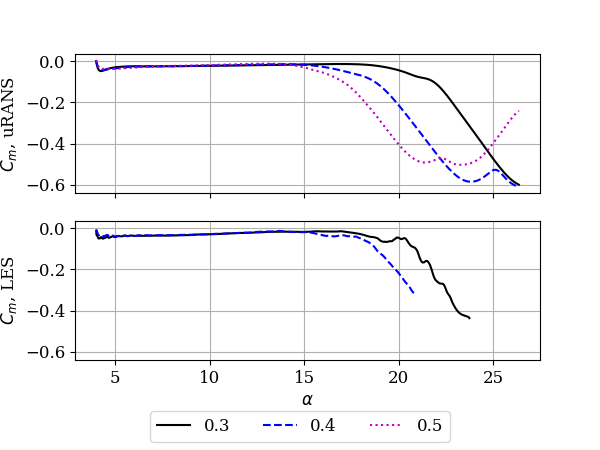}}
    \hspace*{\fill}
    \caption{Comparison of $C_l$ (a) and $C_m$ (b) variation with  $\alpha$ for uRANS and LES (if available), for different $\mach$ values}
    \label{fig:cs_urans_les}
\end{figure}

Figure~\ref{fig:cs_urans_les} presents $C_l$ (panel a) and $C_m$ (panel b) variation with $\alpha$ for uRANS and available LES data.
The top panels correspond to uRANS data and the bottom panels correspond to LES data.
From the curves for $\mach = 0.3$, the peak $C_l$ location and sharp divergence in $C_m$ are delayed in uRANS compared to LES.
This trend has also been observed in our prior work at  $Re \sim 2 \times 10^5$ \citep{Sudharsan2023}.

%%%%%%%%%%%%%%%%%%%%%%%%%%%%%%%%%%

\subsection{Results for \texorpdfstring{$M_\infty = 0.3$}{M = 0.3}}

\subsubsection{Flow field illustration}
We first present a detailed discussion of flow evolution for the lower end of the Mach number range considered, namely, $\mach=0.3$, where stall occurs through APG-induced flow breakdown at the leading edge, without significant shock interaction.
The sequence of flow events is illustrated through space-time contours of the pressure coefficient, $C_p$, and the skin friction coefficient, $C_f$, over the suction surface.
The abscissa corresponds to the normalized chord-wise distance, while the ordinate corresponds to increasing $\alpha$. 
The increase in peak suction near the leading edge is observed from contours of negative $C_p$.
When the APG following the suction peak reaches a certain magnitude, leading edge flow breaks down, the suction peak collapses ($\alpha \sim 17.3^{\circ}$), and shed vorticity organizes itself into a coherent DSV.
The DSV propagates downstream (surface imprint seen in contours of $C_p$ and $C_f$) as $\alpha$ continues to increase, leading to an increase in lift coefficient, even as the peak suction continues to decrease.
There is some trailing-edge separated flow that propagates upstream, however, even at its largest extent, it does not extend upstream of $0.7c$ and there is no interaction with the separated flow at the leading edge.
The space-time diagram for $\mach = 0.4$ remains qualitatively similar, with earlier suction collapse and DSV formation and is not shown here.
\begin{figure}[htb!]
    \centering
    \hspace*{\fill}
    \subcaptionbox{$-C_p$\label{fig:m0p3_xt_cp_re1m}}{\incfig[width=0.45\textwidth]{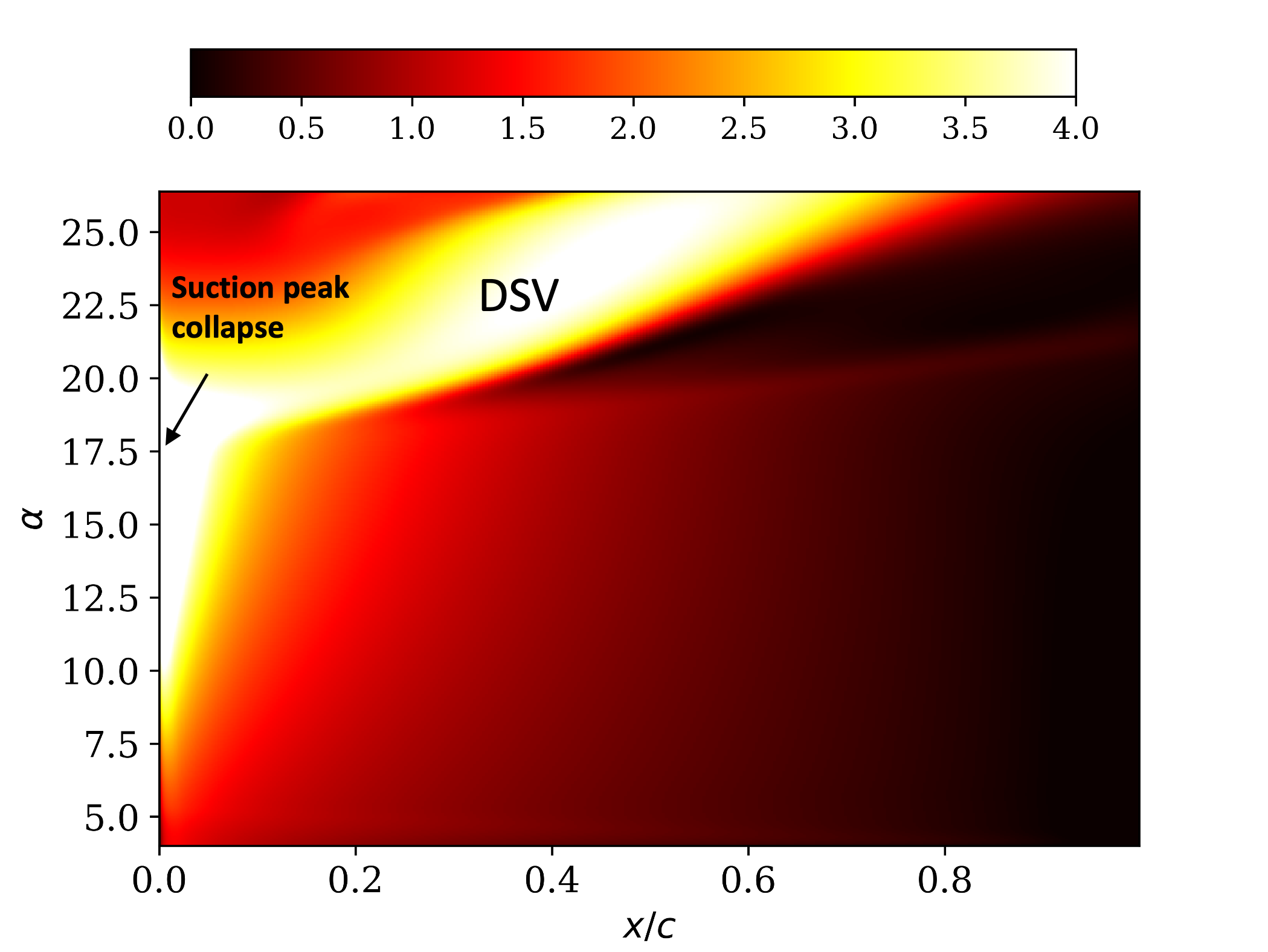}}
    \hfill
    \subcaptionbox{$C_f$\label{fig:m0p3_xt_cf_re1m}}{\incfig[width=0.45\textwidth]{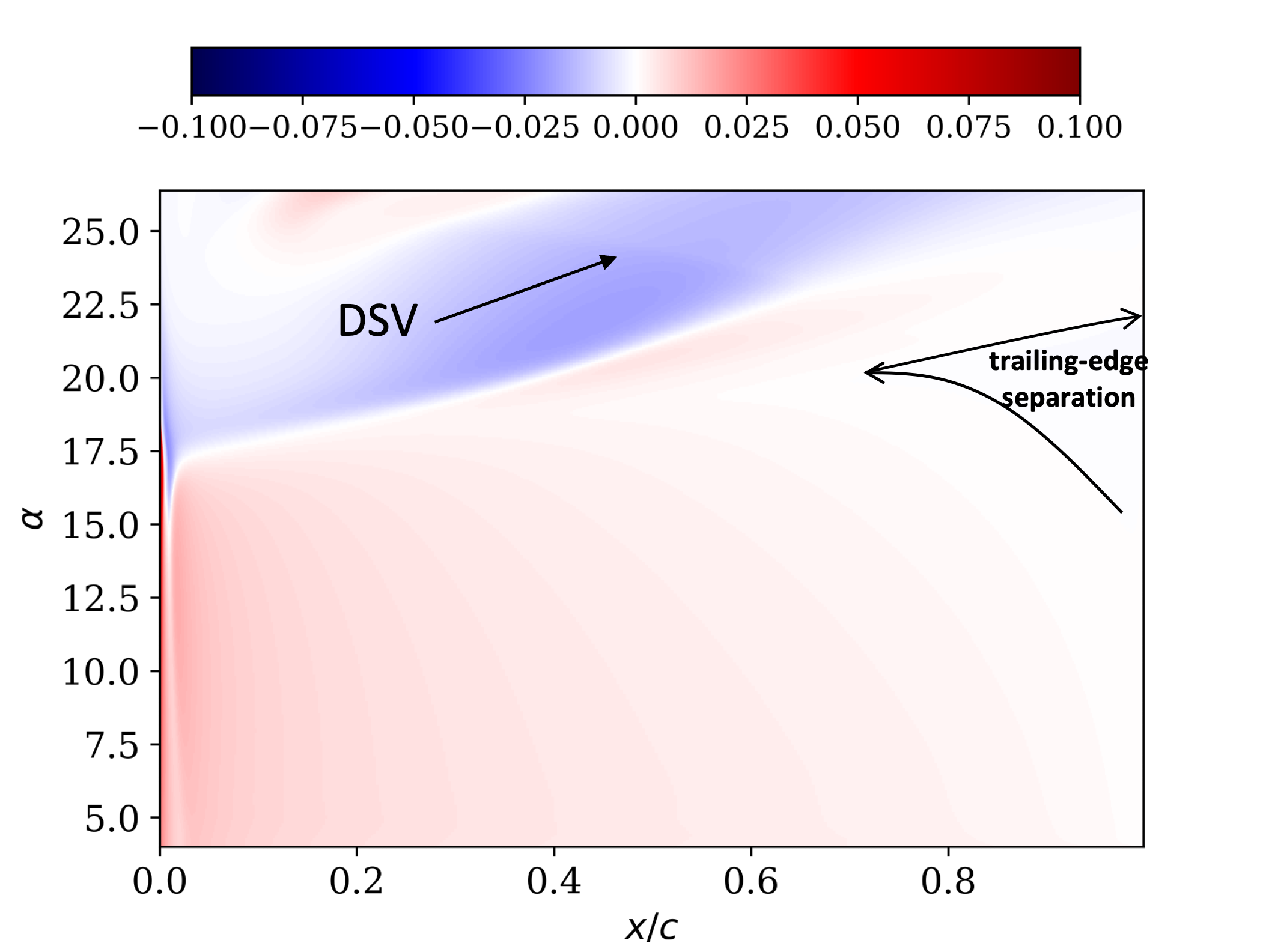}}
    \hspace*{\fill}
    \caption{Space-time contours for $M_{\infty} = 0.3$.
    The ordinate is angle of attack ($\alpha$), which corresponds to time for a constant-rate pitch-up motion.
    \label{fig:m0p3_xt_re1m}}
\end{figure}

\begin{figure}[htb!]
    \centering
    \hspace*{\fill}
        \subcaptionbox{$\alpha = \ang{14.4}$\label{fig:m0p3_lesupersonic}}{\incfig[width=0.45\textwidth]{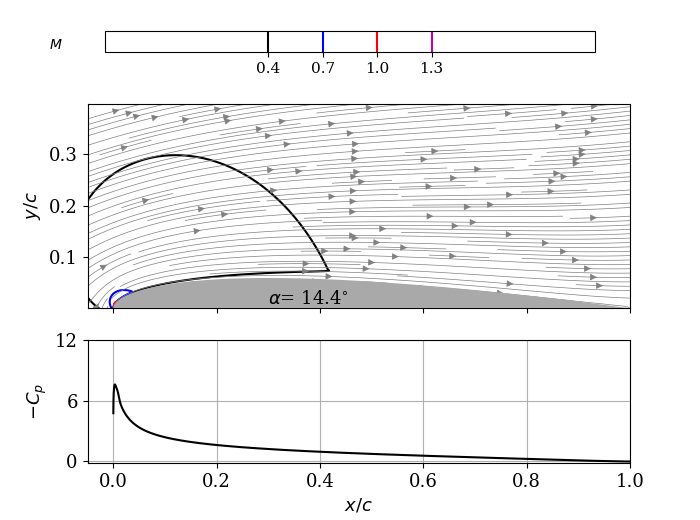}}
    \hfill
        \subcaptionbox{$\alpha = \ang{17.3}$\label{fig:m0p3_DSV}}{\incfig[width=0.45\textwidth]{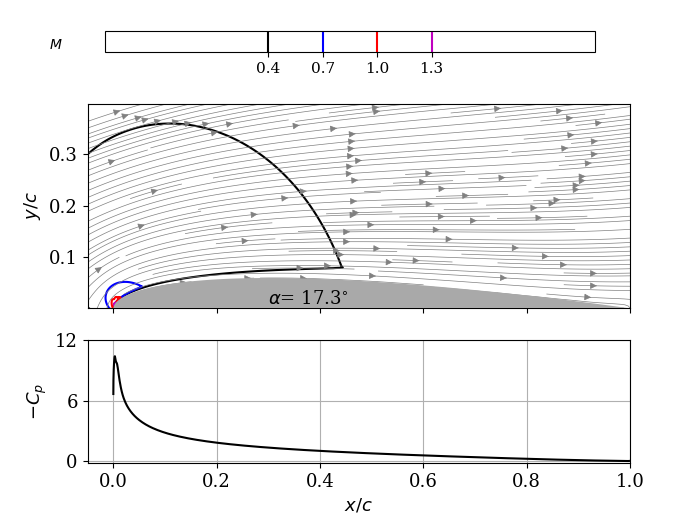}}
    \hspace*{\fill}
	\caption{(a) Supersonic flow above the shear layer at the leading edge, and (b) subsonic DSV formation.
    The top panel shows streamlines overlaid with contours of local Mach number and the bottom panel shows the distribution of $-C_p$ over the airfoil suction surface for $\mach=0.3$.
    \label{fig:supersonic_flowviz_re1m}}
\end{figure}

\Cref{fig:m0p3_lesupersonic,fig:m0p3_DSV} show the flow fields corresponding to the first appearance of supersonic flow near the leading edge and the initial stages of DSV formation, respectively, for $\mach = 0.3$.
The top panels of the figures show streamlines overlaid with local flow Mach number contours and the bottom panel shows the distribution of $C_p$ over the suction surface.
As seen from \cref{fig:m0p3_lesupersonic}, a small pocket of supersonic flow (in red) occurs around $\alpha \sim \ang{14.4}$, very close to the leading edge.
The flow downstream remains fully subsonic, and as seen from panel b, the DSV develops in fully subsonic flow at $\alpha \sim \ang{17}$.
Note the lack of a large suction peak corresponding to the vortex core.

\begin{figure}[htb!]
    \centering
    \hspace*{\fill}
        \subcaptionbox{$\alpha = \ang{14.4}$\label{fig:m0p3_event1_shock}}{\incfig[width=0.45\textwidth]{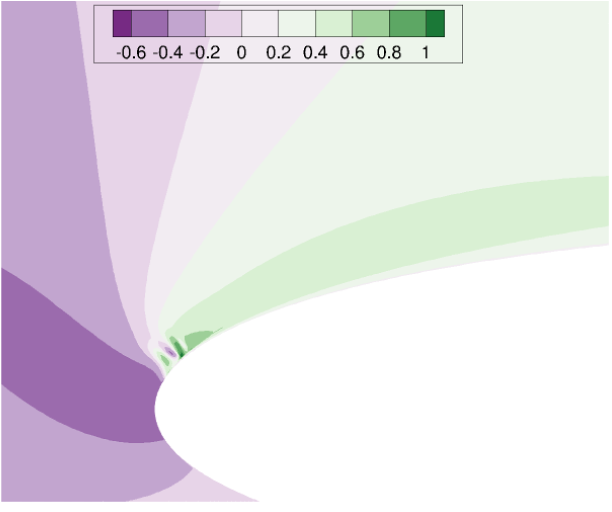}}
    \hfill
        \subcaptionbox{$\alpha = \ang{17.3}$\label{fig:m0p3_event2_shock}}{\incfig[width=0.45\textwidth]{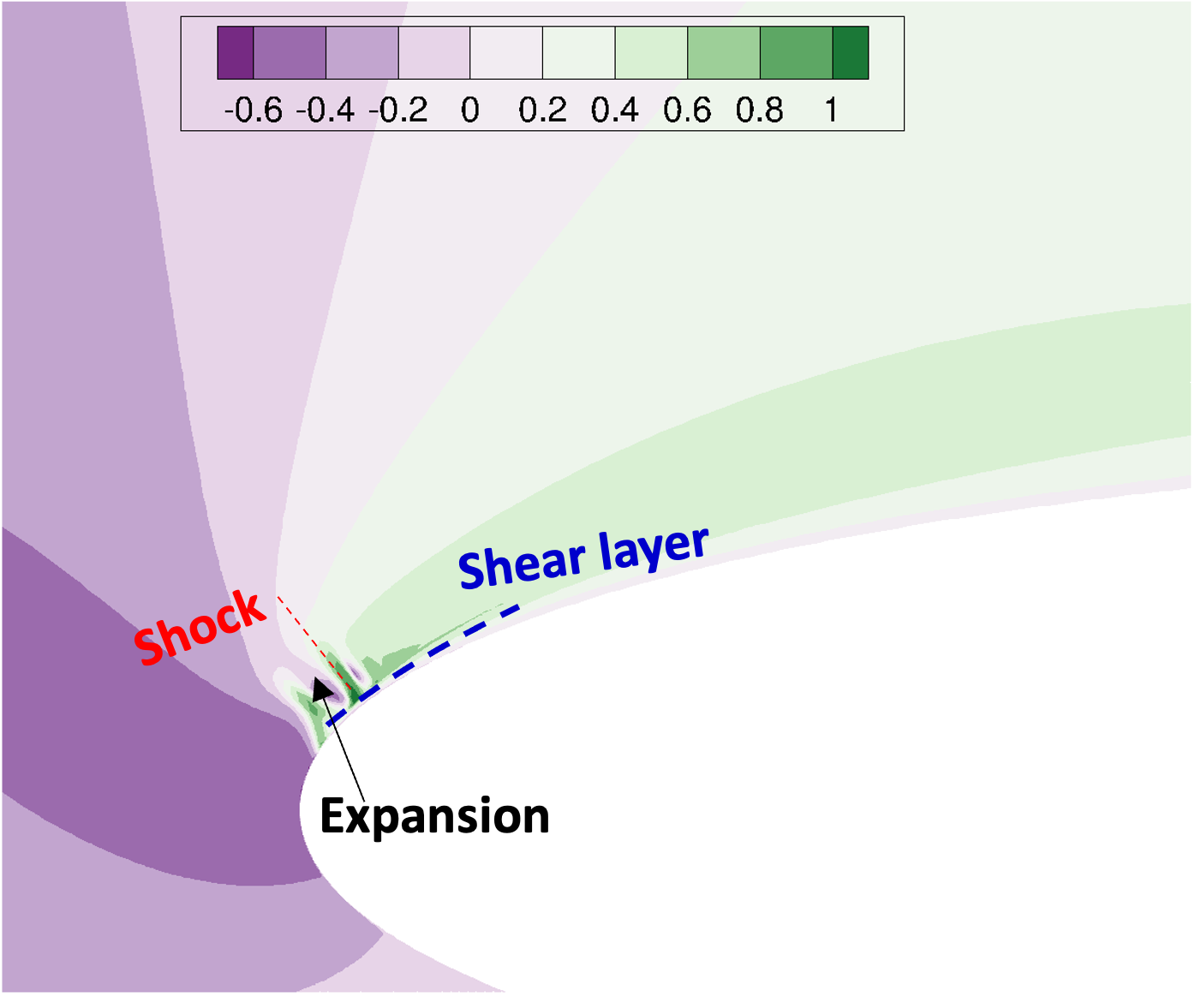}}
    \hspace*{\fill}
	\caption{Shock contours ($(\vb{U} \cdot \grad p )/(a \, \abs{\grad p})$) over the airfoil (a) immediately after the first occurrence, and (b) at the instance of leading-edge suction collapse, for $\mach = 0.3$
    \label{fig:m0p5_shock_re1m}}
\end{figure}

\Cref{fig:m0p3_event1_shock,fig:m0p3_event2_shock} show the shock and expansion structures around the airfoil leading edge corresponding to $\ang{14.4}$ and  $\ang{17.3}$, for $\mach = 0.3$.
Contours of $(\vb{U} \cdot \grad p )/(a \, \abs{\grad p})$
are shown, where $\mathbf{U}$, $a$ and $p$ are local values of the velocity vector, sound speed and pressure, respectively, in the airfoil frame of reference.
This quantity represents the local flow Mach number along the direction of the local pressure gradient and can be used to identify normal shocks.
The two instances shown are at $\alpha = \ang{14.4}$, just after the shock waves first appear, and at $\ang{17.3}$, corresponding to maximum peak suction.
Panel b of the figure shows the development of normal shock waves close to the airfoil leading edge.
For $\mach = 0.4$ (not shown), stronger shock structures that have a larger extent in the streamwise and stream-normal directions are observed, with the general progression of events remaining the same as $\mach = 0.3$.

\subsubsection{Leading-edge stall criteria}

\begin{figure}[htb!]
    \centering
    \hspace*{\fill}
        \subcaptionbox{$LESP/LESP_{\max}$\label{fig:lesp_all}}{\incfig[width=0.48\textwidth]{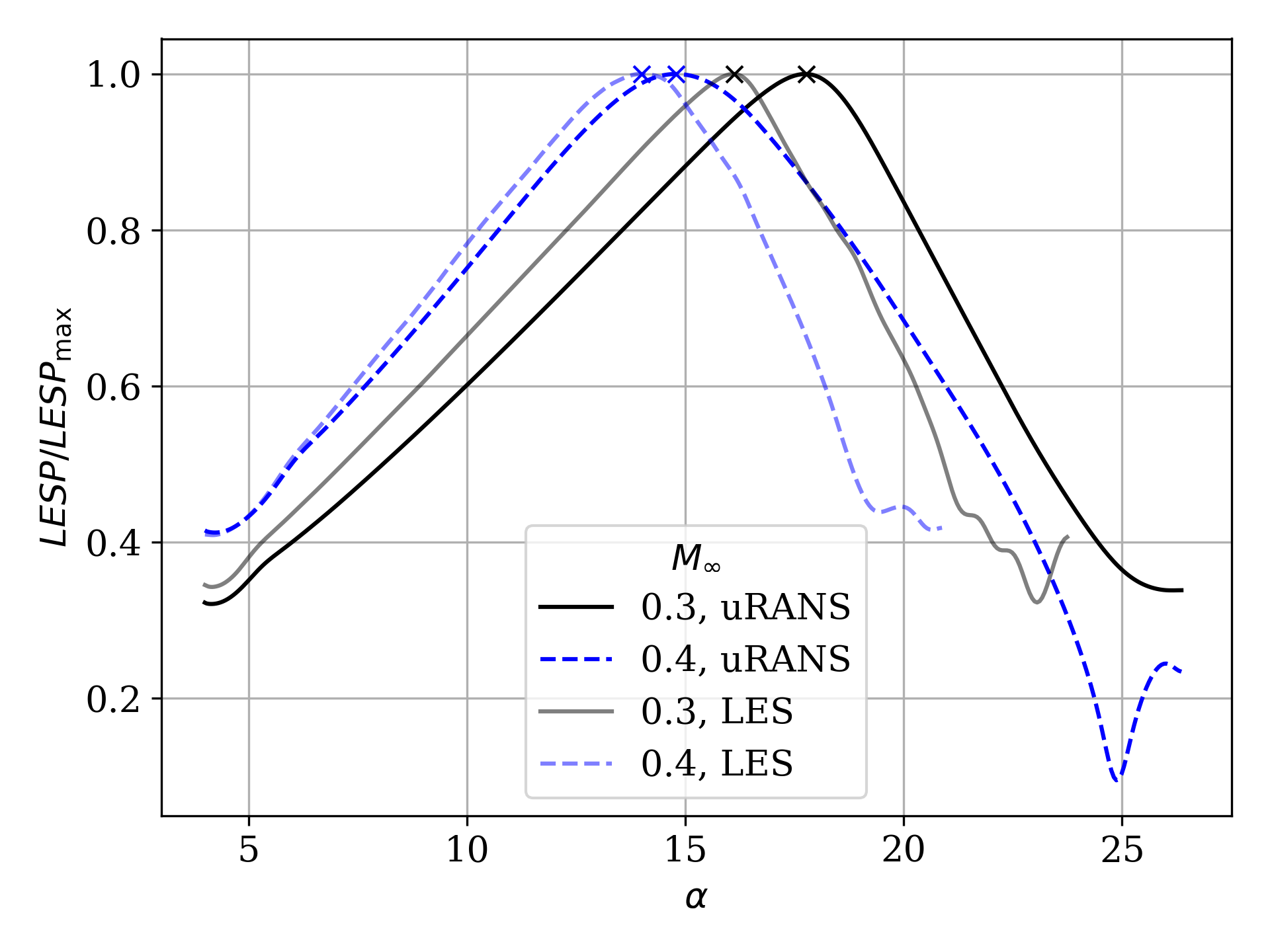}}
    \hfill
        \subcaptionbox{$\abs{BEF}/\abs{BEF}_{\max}$\label{fig:bef_all}}{\incfig[width=0.48\textwidth]{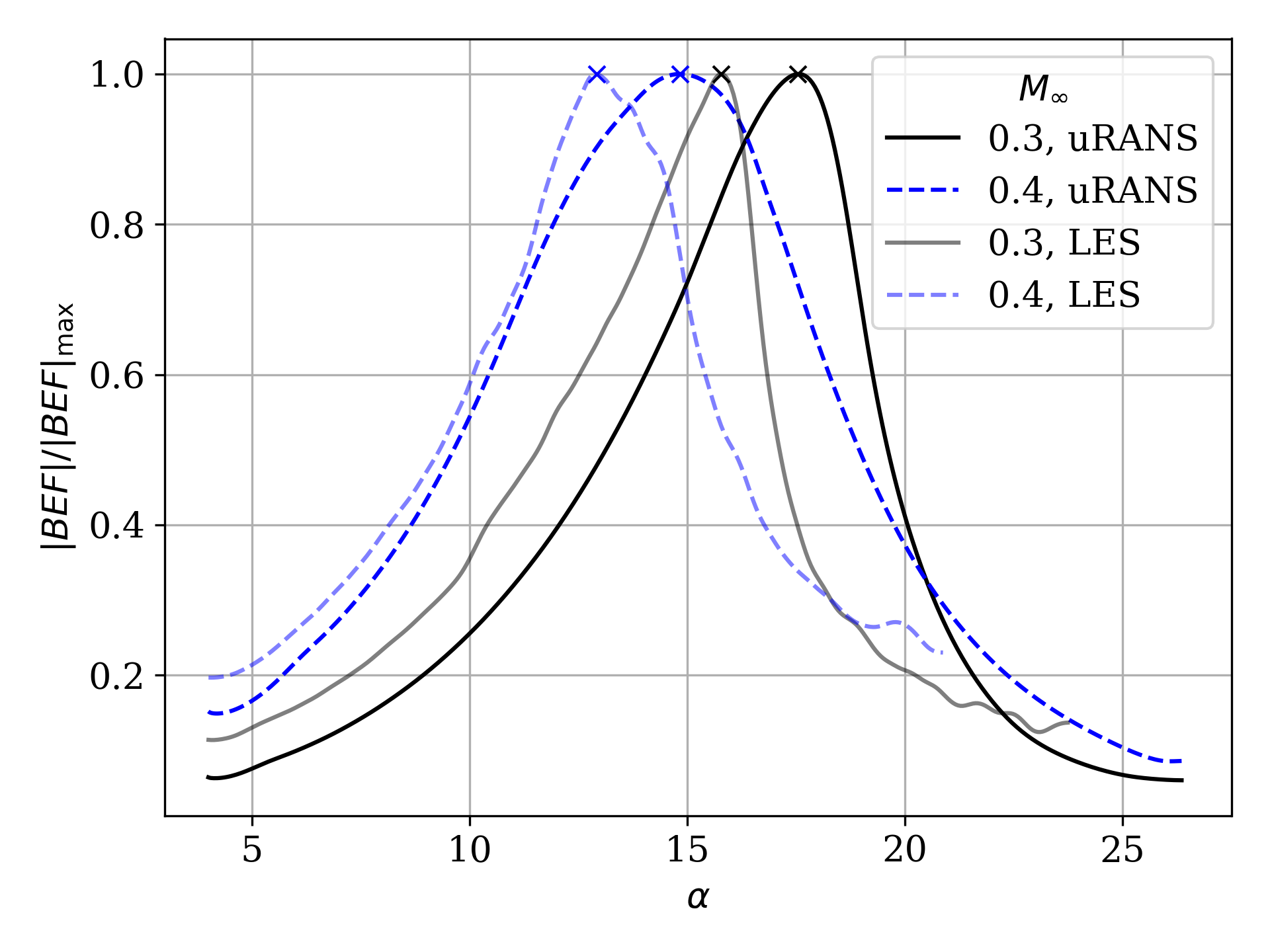}}
    \hspace*{\fill}
    \caption{Variation of the leading-edge stall criteria, $LESP$ and $BEF$, with $\alpha$.
    \label{fig:lesp_bef_all}}
\end{figure}

\Cref{fig:lesp_bef_all} presents the variation with $\alpha$ of $LESP$ (panel a) and $BEF$ (panel b) for $\mach = 0.3$ and  $0.4$, for both uRANS and LES. 
The criteria have been normalized by their maximum magnitudes for consistent comparisons across cases.
The stall criteria demonstrate critical behavior around stall onset by reaching their maximum magnitudes.
As pointed out earlier, lift stall and other events occur earlier for the LES cases, resulting in the $LESP$ and $BEF$ profiles peaking earlier.
While the peaks occur later for the uRANS cases, DSV formation, lift stall, and other events also occur later.
Therefore, the criteria effectively indicate stall onset in advance of DSV formation.
The instance of $\max(\abs{BEF})$ occurs exactly or immediately following peak suction collapse for all the cases.
The reason for the $BEF$, an integrated quantity, closely mimicking the peak suction, a point quantity, is explained in detail in \cite{Sudharsan2022}.
DSV formation occurs between \qtyrange{0.2}{0.5}{\degree} after the stall criteria are reached, as found from examining $C_p$ distributions and streamlines over the airfoil.
Therefore, based on the current results, these criteria are suitable for direct application in the mild to moderately compressible regimes.
These cases do not have significant shock interactions with the boundary layer flow.
To examine the effectiveness of the criteria when shock interactions are more pronounced, we examine the results for $\mach = 0.5$ next.

\subsection{Results for \texorpdfstring{$M_\infty = 0.5$}{M = 0.5}}
\subsubsection{Flow field illustration}
\begin{figure}[htb!]
    \centering
    \hspace*{\fill}
    \subcaptionbox{$-C_p$\label{fig:m0p5_xt_cp_re1m}}{\incfig[width=0.48\textwidth]{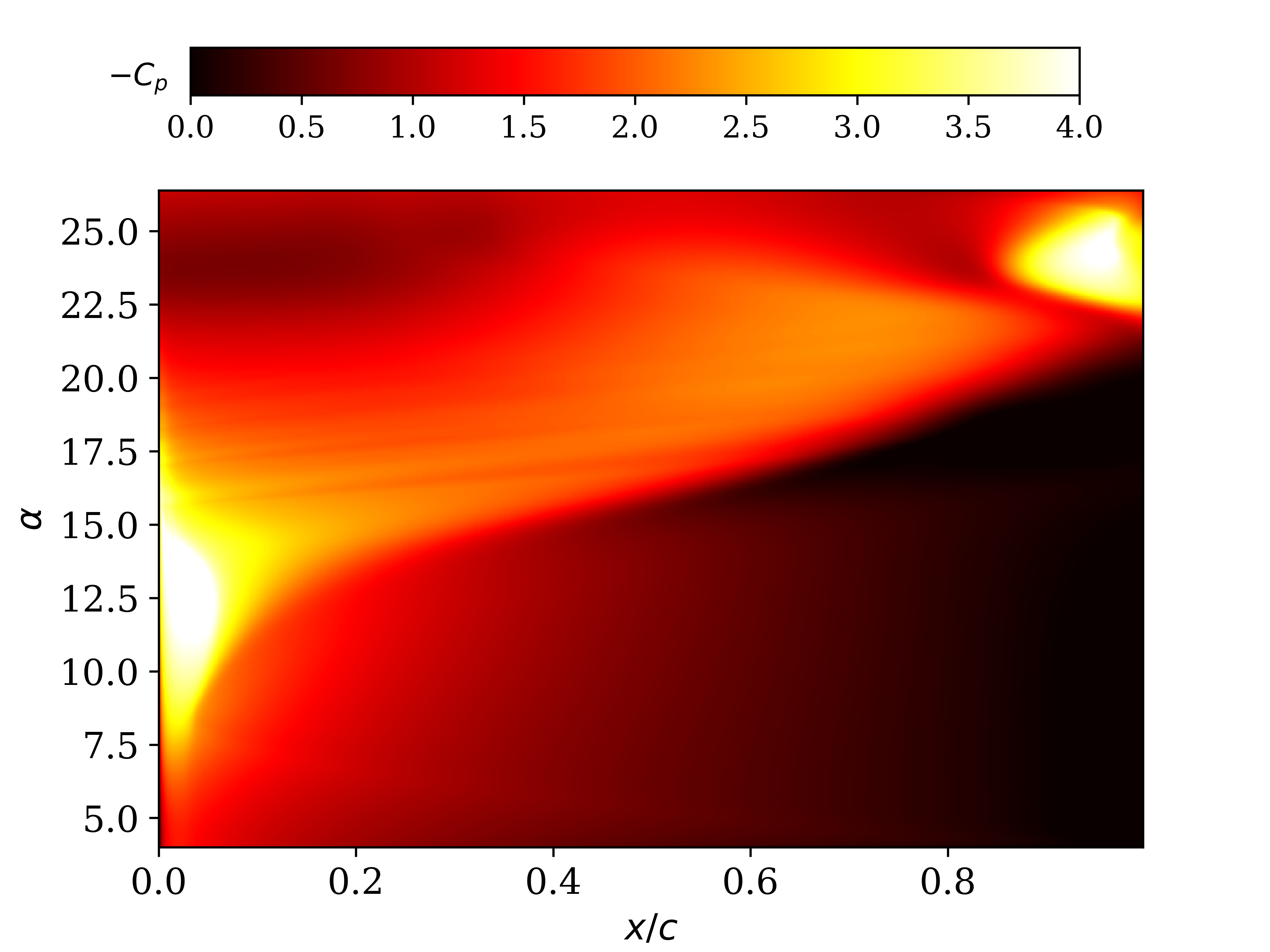}}
    \hfill
    \subcaptionbox{$C_f$\label{fig:m0p5_xt_cf_1m}}{\incfig[width=0.48\textwidth]{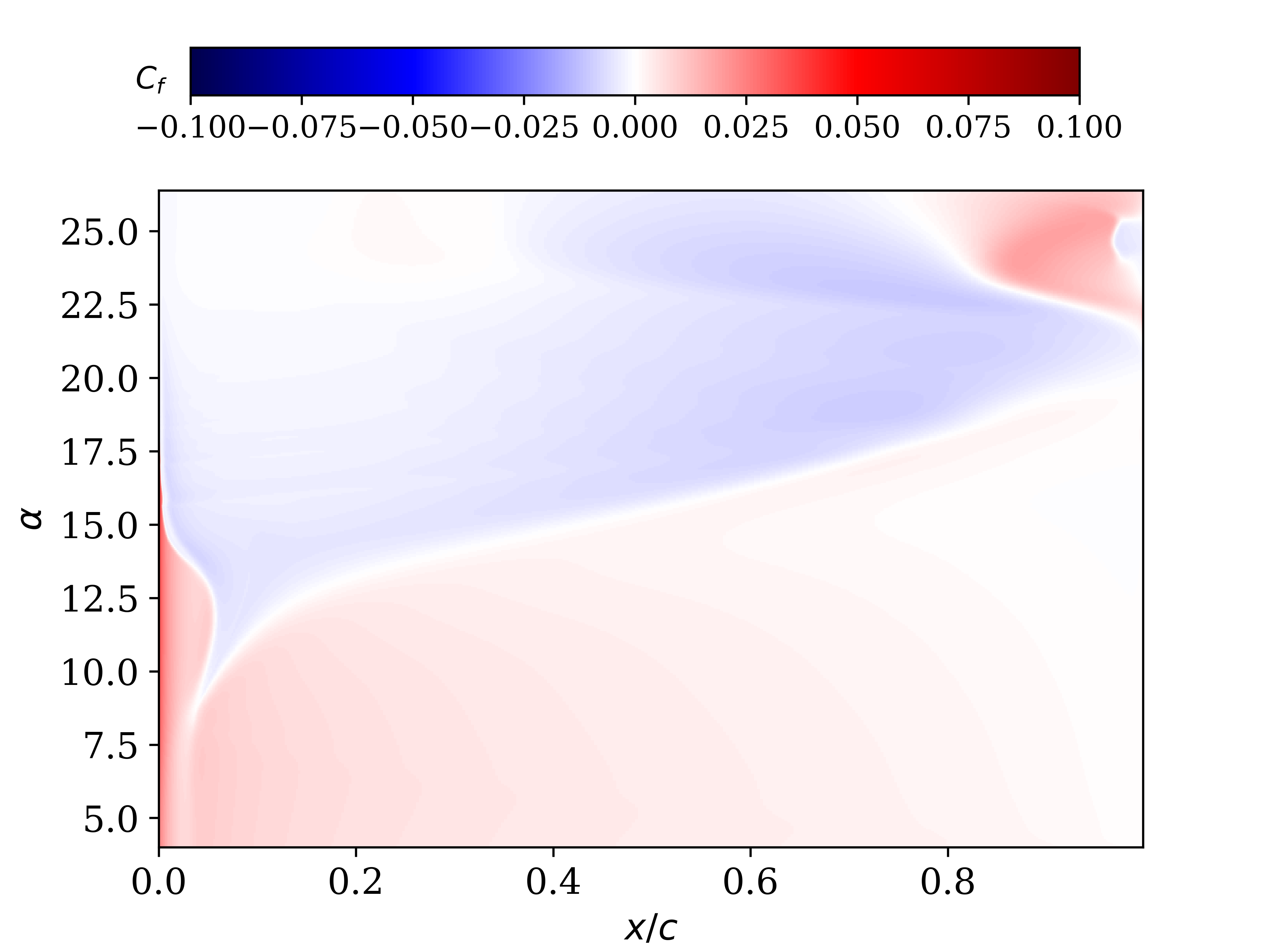}}
    \hspace*{\fill}
    \caption{Space-time contours for the case with $M_{\infty} = 0.5$ and $Re=1 \times 10^6$.
    \label{fig:m0p5_xt_1m}}
\end{figure}

\begin{figure}[htb!]
    \centering
    \hspace*{\fill}
    \subcaptionbox{$\alpha = \ang{7.1}$\label{fig:m0p5_lesupersonic}}{\incfig[width=0.45\textwidth]{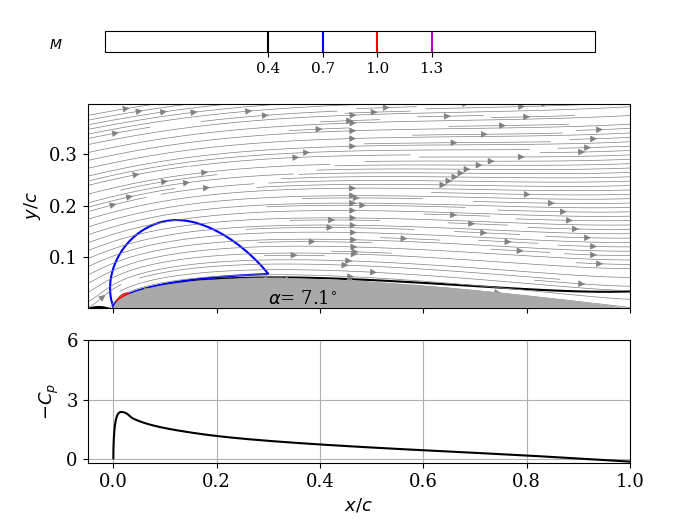}}
    \hfill
    \subcaptionbox{$\alpha = \ang{13.7}$\label{fig:m0p5_DSV}}{\incfig[width=0.45\textwidth]{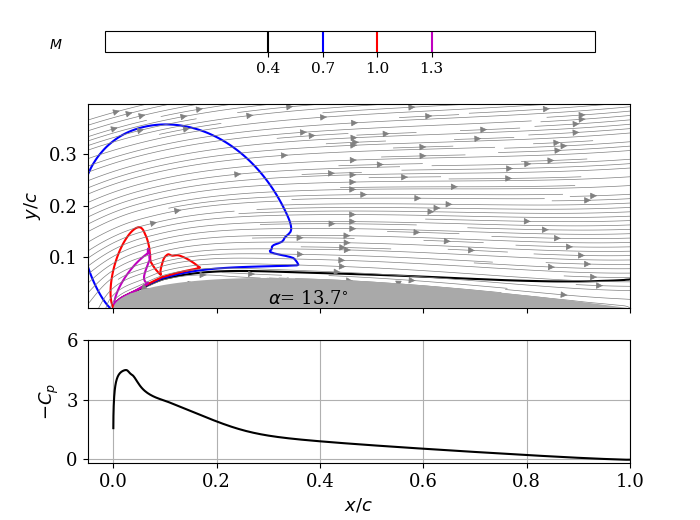}}
    \hspace*{\fill}
	\caption{(a) Supersonic flow above the shear layer at the leading edge, and (b) maximum peak suction over the leading edge.
\label{fig:supersonic_flowviz_re1m_m0p5}}
\end{figure}

Space-time contours of $C_p$ and $C_f$ for the $M_\infty = 0.5$ case are shown in \cref{fig:m0p5_xt_1m}.
The flow close to the leading edge reaches supersonic speeds at a much lower $\alpha$, around $\ang{7}$.
As with previous cases, a separated shear layer exists downstream of this supersonic region.
However, the extent of the supersonic flow region and strength of the shock structures that are formed, are dramatically higher compared to the lower $\mach$ cases.
The extent of suction near the leading edge is more spread out in the chordwise direction.
As the airfoil pitches up, a series of shock waves form within the supersonic region and propagate downstream.
A weak DSV begins to roll up around $\ang{12.5}$ within the shear layer downstream of the supersonic flow region. 
Leading edge suction collapses around $\ang{13.7}$ and following this point, the shock and rarefaction structures move upstream as the DSV grows in size.
The top panel in \cref{fig:supersonic_flowviz_re1m_m0p5} shows streamlines overlaid with local flow Mach number contours and the bottom panel shows the distribution of $C_p$ over the suction surface.
The two instances shown are at $\alpha = \ang{7.1}$, just after the flow first reaches supersonic speeds, and at $\ang{13.7}$, when the peak suction collapses.
The DSV is formed in a fully subsonic flow region (\cref{fig:m0p5_DSV}) and grows in size.
As the separation point of the shear layer moves towards the leading edge and the acceleration over the airfoil reduces, the supersonic pockets reduce in extent and disappear.
Note the lack of a large suction peak corresponding to the vortex core.
As the DSV is shed and $\abs{C_p}$ drops over most of the airfoil suction surface, a strong counter-clockwise vortex rolls up around the trailing-edge (as pointed out in \cref{fig:m0p5_xt_1m}).
The progression of events remains similar as for $\mach=0.4$, though the extent of the supersonic region is larger in the streamwise and stream-normal directions.

\begin{figure}[htb!]
    \centering
    \hspace*{\fill}
    \subcaptionbox{$\alpha = \ang{7.1}$\label{fig:m0p5_event1_shock}}{\incfig[width=0.45\textwidth]{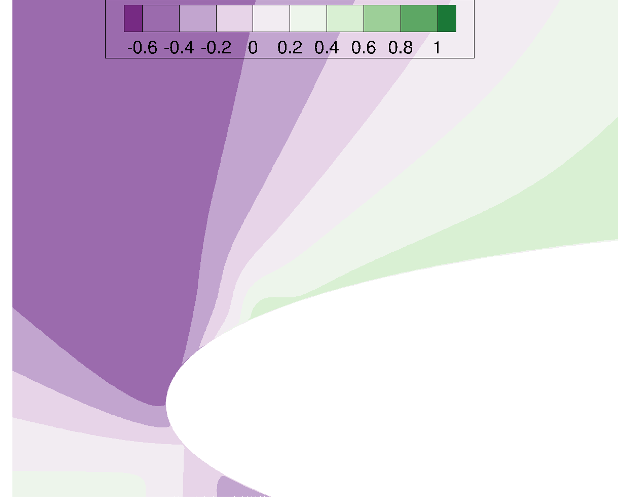}}
    \hfill
    \subcaptionbox{$\alpha = \ang{13.7}$\label{fig:m0p5_event2_shock}}{\incfig[width=0.45\textwidth]{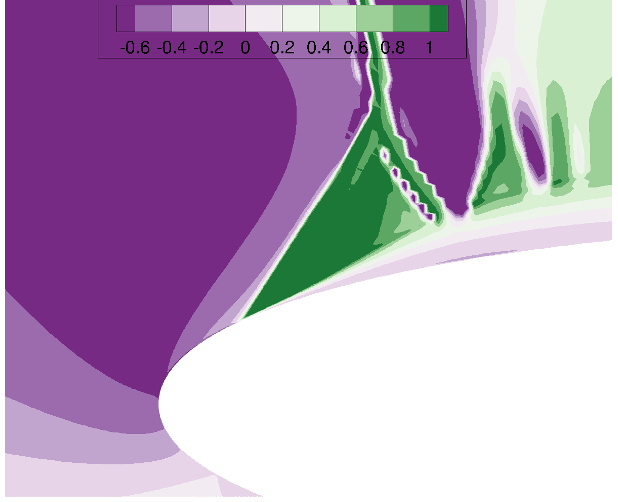}}
    \hspace*{\fill}
	\caption{Shock contours ($(\mathbf{U} \cdot \bm{\nabla} p )/(a \, \abs{\bm{\nabla} p})$) over the airfoil (a) immediately after the first occurrence, and (b) at the instance of leading-edge suction collapse.
\label{fig:m0p5_shock}}
\end{figure}

\Cref{fig:m0p5_shock} shows the rarefaction and shock waves that form near the leading edge above the separated shear layer as the airfoil pitches up.
Contours of $(\vb{U} \cdot \grad p )/(a \, \abs{\grad p})$ are shown at the same two instances as the previous figure.
The development of strong shocks is seen in panel b of \cref{fig:m0p5_shock} as flow acceleration around the leading edge continues to increase with increasing $\alpha$.
Higher-fidelity techniques, such as LES, can help better illuminate the interactions between the shock waves and the shear layer.

\subsubsection{Leading-edge stall criteria}
Next, we compare the variation of the two leading-edge criteria, $\max(LESP)$ and $\max(\abs{BEF})$.
Similar to the behavior of $\max(\abs{C_p})$ (see bottom panel of \Cref{fig:aerodyn_coeff_all_re1m}), $\abs{BEF}$ demonstrates a broad peak resulting from the suction peak location being downstream of a large supersonic flow region.
$\max(LESP)$ demonstrates a sharp peak; however, it occurs after DSV formation.
Though the $\abs{BEF}$ begins to flatten out shortly after the $LESP$ peak, it does not demonstrate a clear peak as in the prior cases. 
The disparity between the instance of DSV formation and the criteria being reached increases with stronger shock interaction effects at this $Re$.
Therefore, the definition of $LESP$ and  $BEF$ might need to be revisited for effective stall indication in moderately compressible regimes.

\begin{figure}[htb!]
    \centering
    \hspace*{\fill}
    \subcaptionbox{$LESP/LESP_{\max}$\label{fig:lesp_m0p5}}{\incfig[width=0.48\textwidth]{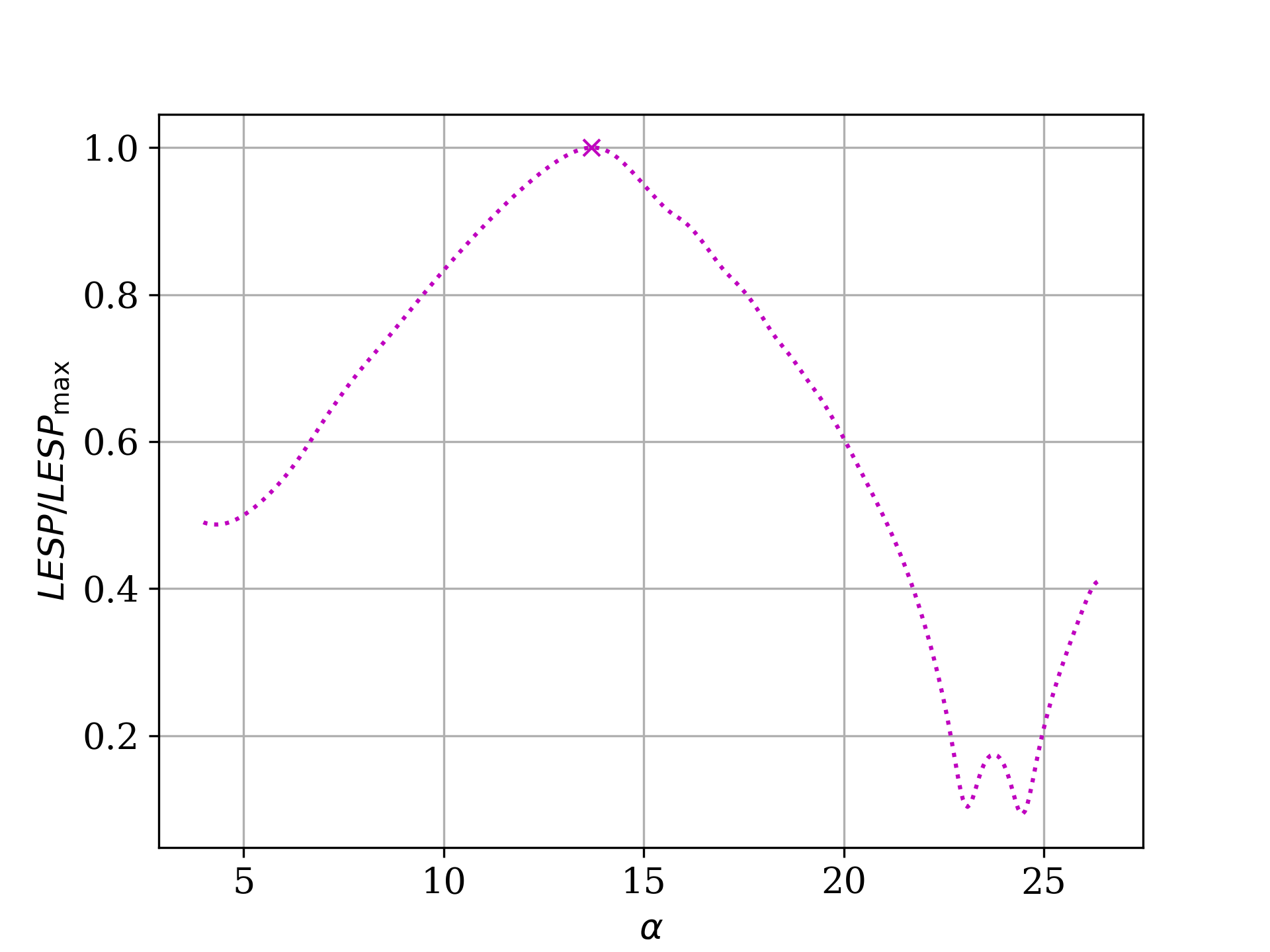}}
    \hfill
    \subcaptionbox{$\abs{BEF}/\abs{BEF}_{\max}$\label{fig:bef_m0p5}}{\incfig[width=0.48\textwidth]{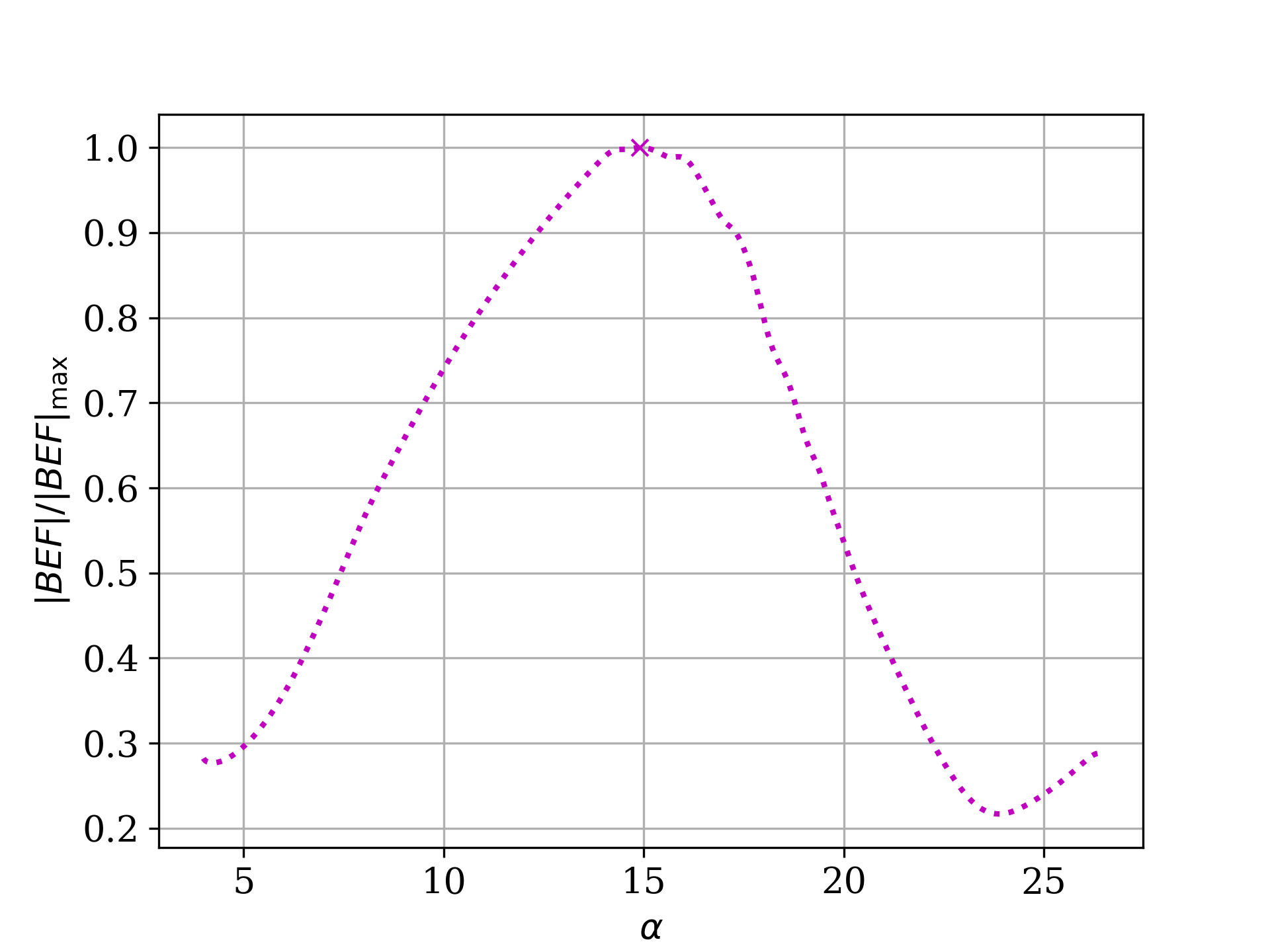}}
    \hspace*{\fill}
    \caption{Variation with $\alpha$ of the leading-edge stall criteria, $LESP$ and $BEF$ (normalized by their respective maximum magnitudes), for $\mach = 0.5$.
    \label{fig:lesp_bef_m0p5}}
\end{figure}

\section{Conclusion}
\label{sec:conclusion}

The effectiveness of leading-edge dynamic stall criteria (max($|BEF|$) and max($LESP$)) in a moderately compressible regime was considered in the present work.
These criteria have been effective in indicating stall onset ahead of DSV formation in the incompressible regime and mildly compressible regime.
In the present work, we extend their application to $Re = 1 \times 10^6$ and $M_{\infty}$ between 0.3 - 0.5.
While compressibility effects tend to promote APG-induced stall, shock-induced separation effects present a challenge for such leading edge stall criteria.
Based on uRANS simulations, we observe that the instance of DSV formation falls very close to the instance of the criteria being reached.
For the highest $\mach$ number case, DSV formation occurs before the criteria are reached. 
These results suggest that the definitions of these stall criteria need to be revisited in the presence of stronger compressibility effects where shock-induced separation may play a role in the stall process. 
Furthermore, employing higher-fidelity techniques such as LES for $M_{\infty} \ge 0.5$ would better validate these results and facilitate their effective application in the presence of strong compressibility effects.
These approaches could be considered in future studies.

\section*{Funding Sources}
Funding for this research is provided by the National Science Foundation (grants CBET-1554196 and 1935255)
and the US Air Force Office of Scientific Research (Award \# FA9550-23-1-0016).
Co-author Sharma acknowledges the support provided to him by the 2021 AFOSR Summer Faculty Fellowship.

\section*{Acknowledgements}
We acknowledge the computational resources provided by Iowa State University.

%%%%%%%%%%%%%%%%%%%%%%%%%%%%%%%%%
\section*{Declaration of Interests}
The authors report no conflict of interest.

%%%%%%%%%%%%%%%%%%%%%%%%%%%%%%%%%
\bibliography{main}
\bibliographystyle{unsrtnat}

\end{document}